\documentclass[twocolumn,10pt,tran,compsoc]{revtex4-1}
\usepackage{amssymb}
\usepackage{graphicx}
\usepackage{natbib} 
\usepackage{mathptmx}
\usepackage{comment}
\begin{document}

\title{Scaling in Modeling of Core Losses in Soft Magnetic Materials Exposed to Nonsinusoidal
Flux Waveforms and DC Bias Condition}   

\author{Adam Ruszczyk}
\email[e-mail: ]{adam.ruszczyk@pl.abb.com}
\affiliation{ ABB Corporate Research, Starowi\'{s}lna 13a, 31-038 Krak\'{o}w, Poland}
\author{Krzysztof Sokalski}%
\email[e-mail: ]{sokalski\_krzysztof@o2.pl}
\affiliation{Institute of Computer Science, 
Cz\c{e}stochowa University of Technology,
Al. Armii Krajowej 17, 42-200 Cz\c{e}stochowa, Poland}
\author{Jan Szczyg{\l}owski}
\email[e-mail: ]{jszczyg@el.pcz.czest.pl}
\affiliation{Institute of Power Engineering, Cz\c{e}stochowa University of Technology, Al. Armii Krajowej 17, 42-200 Cz\c{e}stochowa, Poland}

\begin{abstract}
Assuming that core loss data of Soft Magnetic Materials obey scaling relations, models describing the power losses in materials exposed to nonsinusoidal flux waveforms and DC Bias conditions have been derived.
In order to test these models, the measurement data for two materials have been collected and  the core losses calculated. Agreement between the experimental data and the model predictions is satisfactory.

\end{abstract}
\maketitle

 \section*{Introduction}\label{I}
The usual waveforms in power electronics are square
waves or a superposition of square waves rather than sine waves.  Some magnetic
components also operate under a DC bias, which has a significant influence on the core losses. During the last two ldecades  significant progress in modeling the dependences of core losses both 
due to square wave parameters
and DC bias  has been achieved. The derived methods for handling the core losses in soft magnetic materials (SMM) exposed to nonsinusoidal
flux waveforms have been based on the Steinmetz Equation \citep{bib:Steinmetz}. These methods include the Modified Steinmetz Equation (MSE) \citep{bib:Albach}, \citep{bib:Reinert},
Generalized Steinmetz Equation (GSE) \citep{bib:Li}, Improved Generalized Steinmetz Equation (iGSE) \citep{bib:Venka} \citep{bib:Boss1} and 
Improved-improved Generalized Steinmetz Equation (i$^{2}$GSE) \citep{bib:Ecklebe}. Methods and models exist, however, based on the assumption that the shape of
the waveform does not matter and as a result only look at peaks: Scaling of power Losses in SMM \citep{bib:Sokal1},\citep{bib:Sokal2},\citep{bib:Sokal3} and Field-Extrema Hysteresis Model (FHM), \citep{bib:Cale}.
Both do not capture the effects of the waveform. Is this feature a disadvantage or an advantage? All the models we are talking about have been derived for prediction of losses for the designing processes. If one assumes that the waveform does not vary in such a process then the model is universal and can be applied to any waveform possessing a constant period. Moreover,  if experimental data obey a scaling law then the data collapse is possible. In practise the collapse leads to a reduction of the number of independent variables \citep{bib:Sokal1} - \citep{bib:Sokal3} which makes a model or a method much simpler. For these reasons we will apply the scaling theory in order to describe the core losses in soft magnetic materials exposed to any periodic flux waveform.  In order to take into account  DC Bias we apply the method derived by Van den Bossche, et al. \citep{bib:Boss1}, \citep{bib:Boss2}. 
\\
This paper is organized in the following way. In Section I we present the Scaling approach to modeling of core losses for  any periodic flux waveform. In order to take into account  DC bias we combine the scaling with the Van den Bossche method. Section II is devoted to experiment, measurement data, estimation of the model parameters and comparison of theoretical model with the measurement data.  Section III  contains conclusions. 
\section{Approach to scaling of core losses data }\label{II} 
 Instead of an analysis based on Maxwell's equations \citep{bib:GB},\citep{bib:GB2}  we have assumed that the SMM is a complex system in which the function of power losses obeys the scaling law. This assumption leads to the formula for total power loss in the form of general homogenous function \citep{bib:Stan1}, \citep{bib:Stan2}:
 \begin{eqnarray}
\exists\hspace{2mm} a,b,c\in { \mathbf{R}}:\hspace{1mm}\label{gen4}\\
 \forall \lambda\in { \mathbf{R}}^{+}\hspace{2mm}  
P_{tot}(\lambda^{a}f,\lambda^{b}(\triangle B))=\lambda^{c}P_{tot}(f,(\triangle B)),\nonumber
\end{eqnarray}
where $P_{tot}$ - total loss of power's density, $f$ - frequency, $\triangle B$ - peak to peak magnetic induction.
Substituting $\lambda=(\triangle B)^{-\frac{1}{b}}$ we have derived the general form of $P_{tot}$:
\begin{equation}
\label{general}
P_{tot}(f,(\triangle B))=(\triangle B)^{\beta}F\left(\frac{f}{(\triangle B)^{\alpha}}\right),
\end{equation}
where $F(\cdot)$ was an arbitrary function, $\alpha=\frac{a}{b}$, and $\beta=\frac{c}{b}$. This function depends on the features of the phenomena to be described. Since our measurement data are unable to consider quasi-static losses we choose for $F(\cdot)$ the power series as a rough description of the power losses  $P_{tot}$ \citep{bib:Sokal1}-\citep{bib:Sokal3}: 
\begin{eqnarray} 
\frac{P_{tot}(f,\triangle B)}{(\triangle B)^{\beta}}= [\Gamma_{1}\,\frac{f}{(\triangle B)^{\alpha}}+ \Gamma_{2}\,\left(\frac{f}{(\triangle B)^{\alpha}}\right)^2 +\label{eq8}\\ 
\Gamma_{3}\,\left(\frac{f}{(\triangle B)^{\alpha}}\right)^3 +\Gamma_{4}\,\left(\frac{f}{(\triangle B)^{\alpha}}\right)^4+... ]. \nonumber
\end{eqnarray}

 Note that the left hand side of (\ref{eq8}) is a function of the one effective variable $\frac{f}{(\triangle B)^{\alpha}}$. It is easy to recognize in (\ref{eq8}) the hysteresis losses $P_{h}$ and the eddy current losses $P_{c}$ by the powers  $f$ and $f^{2}$, respectively. All higher terms correspond to excess losses $P_{ex}$. The formula (\ref{eq8}) is the mathematical model, ready to apply.
Values of $\alpha$, $\beta$ and amplitudes $\Gamma_{n}$  have been estimated for the ten selected soft magnetic materials \citep{bib:Sokal1}-\citep{bib:Sokal3}. Very recently (\ref{eq8}) has been applied to Soft Magnetic Composities \citep{bib:Slus}. Obtained results confirm the good quality of this model. Unfortunately, the model fails when a DC Bias is present. In this situation we propose to keep (\ref{eq8}) as a limit for $H_{DC}\rightarrow 0$ and add the second term of a similar  form to (\ref{eq8}) with $\Gamma$-s dependent on $H_{DC}$. These we take from Van den Bossche, et al. \citep{bib:Boss1}, \citep{bib:Boss2}. They have proposed mapping  $H_{DC}$ to the corresponding primary magnetization curve which works very well. In the case of a lack of appropriate experimental data it is possible to use a mathematical model for $H_{DC}$. For instance:
\begin{equation}
\label{mag}
H_{DC}\rightarrow tanh(H_{DC}\cdot c),
\end{equation}
or more generally:
\begin{equation}
\label{mag}
H_{DC}\rightarrow \{tanh(H_{DC}\cdot c_{1}),tanh(H_{DC}\cdot c_{2}),\dots,tanh(H_{DC}\cdot c_{p})\},
\end{equation}
where $c_{i}$ are  free parameters to be determined \citep{bib:Boss1}. By this way we derive the final model's form:
\begin{eqnarray}
\frac{P_{tot}(f,\triangle B, H_{DC})}{(\triangle B)^{\beta}}=\Sigma_{i=1}^{5}\Gamma_{i}\left (\frac{f}{(\triangle B)^{\alpha}}\right)^{i\,(1-x)}+\nonumber\\
\Sigma_{i=1}^{3}\Gamma_{i+5}\left (\frac{f}{(\triangle B)^{\alpha}}\right)^{(i+4)\,(1-x)}tanh(H_{DC}\cdot c_{i}) ,\label{eq9}
\end{eqnarray}
where $x$ is a small phenomenological tuning parameter. This parameter is derived in the following way. (\ref{eq8}) arises as the Maclaurin series of (\ref{general}). This is why expansion of (\ref{eq8}) posseses only integer exponents. However, in nature this needs not be so. Therefore by the small $x$ we introduce a natural correction of the series' exponents. \\ Equation (\ref{eq9}) has been successfully applied to SIFERRIT N87, however playing with data for VITROPERM 500F 18K  and model (\ref{eq9}) we have got an expirence which suggested the following modifications: 1) to split the $x$ parameter into  two independent tuning parameters $x$ and $y$, where  $x$ tunes the bias-independent term, whereas $y$ tunes the part describing the influence of the bias on the losses, and 2) to distribute functions (\ref{mag}) along the $H_{DC}$ axis. These requirements lead to the following model:
\begin{equation}
\label{mag1}
H_{DC}\rightarrow \{tanh(H_{DC}\cdot c_{1}-r_{1}),\dots,tanh(H_{DC}\cdot c_{p}-r_{p})\},
\end{equation}
\begin{eqnarray}
\frac{P_{tot}(f,\triangle B, H_{DC})}{(\triangle B)^{\beta}}=\Sigma_{i=1}^{4}\Gamma_{i}\left (\frac{f}{(\triangle B)^{\alpha}}\right)^{i\,(1-x)}+\nonumber\\
\Sigma_{i=0}^{2}\Gamma_{i+5}\left (\frac{f}{(\triangle B)^{\alpha}}\right)^{(i+y)(1-x)}{tanh(H_{DC}\,c_{i+1}-r_{i+1})},\label{eq10}
\end{eqnarray}
where $r_{i}$ describes the position of the $i$-th function (\ref{mag}).

\section{Experimental data and parameters' estimations}\label{II}

The B-H Loop measurements  have been performed for two materials: SIFERRIT N87 and VITROPERM 500F 18K. The Core Losses per unit volum have been calculated  
as the enclosed area of the B-H loop, multiplied by the frequency $f$. The following factors influence the accuracy of measurements: 1) Phase Shift Error of Voltage and Current $\le 4\%$,  
2) Equipment Accuracy $\le 5,6\% $, 3) Capacitive Couplings $negligible$ (capacitive currents are
relatively lower compared to inductive currents), and  4) Temperature $\le 4\% $. For details of the applied measurement method and the errors of factors  we refer to \citep{bib:Ecklebe},\citep{bib:Ecklebe1}. In order to enable readers to compare their own results with this work we enclose measurement data for SIFERRIT N87 (TABLE \ref{APPENDIX}).\\
The parameter values of (\ref{eq8}) have been estimated for each sample's measurement data by minimization of $\chi^2$ using the Simplex method of Nelder and Mead \citep{recipes}, 
see TABLE \ref{Tab1} and TABLE \ref{Tab2}.

\begin{figure}[t!]
\includegraphics [angle=0, width=8 cm]{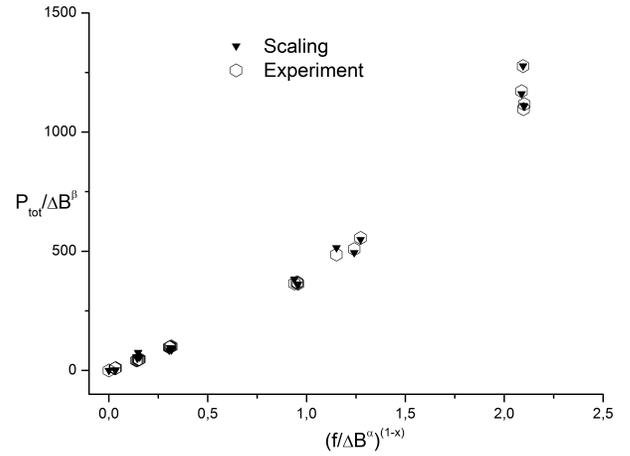}
\caption{  SIFERRIT N87. Projection of measurement points and  scaling theory points (\ref{eq9}) in $((f/ (\triangle B)^{\alpha})^{1-x},P_{tot}/(\triangle B)^{\beta})$ plane.}
{\label{Fig.3a}}
\end{figure}
\begin{figure}[t!]
\includegraphics [angle=0, width=8 cm]{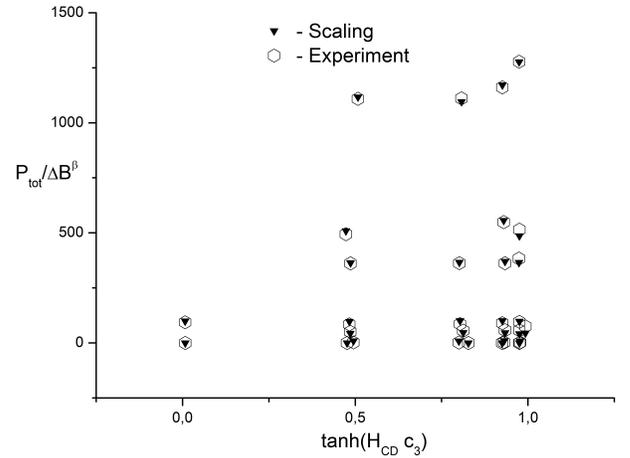}
\caption{SIFERRIT N87. Projection of measurement points and  scaling theory points (\ref{eq9}) in $(tanh(H_{DC} \cdot c_{3}),P_{tot}/(\triangle B)^{\beta})$ plane. }
{\label{Fig.3b}}
\end{figure}

\begin{figure}[t!]
\includegraphics [angle=0, width=8 cm]{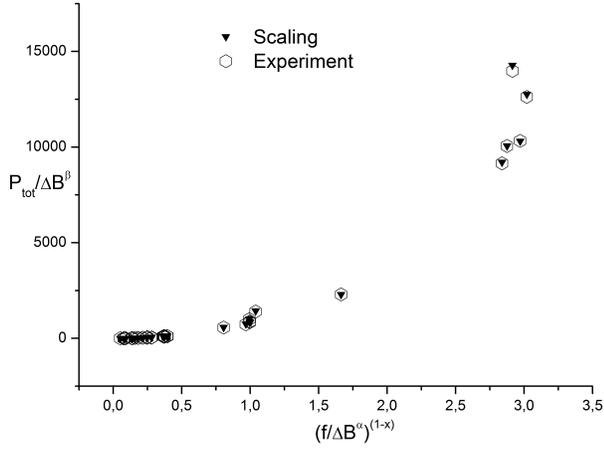}
\caption{ VITROPERM500F 18K. Projection of measurement points and  scaling theory points (\ref{eq10}) in $((f/ (\triangle B)^{\alpha})^{1-x},P_{tot}/(\triangle B)^{\beta})$ plane.}
{\label{Fig.4a}}
\end{figure}
\begin{figure}[t!]
\includegraphics [angle=0, width=8cm]{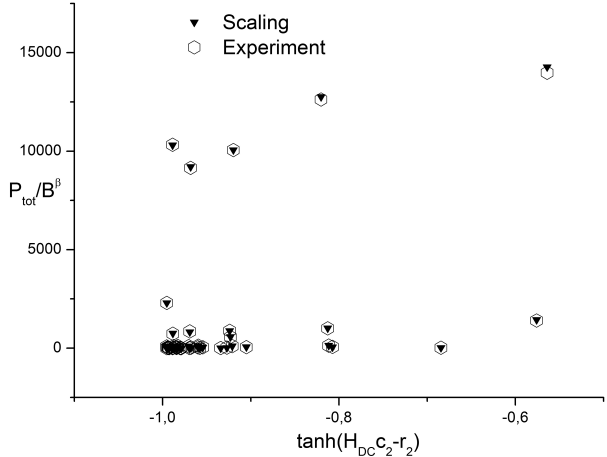}
\caption{VITROPERM500F 18K. Projection of measurement points and  scaling theory points (\ref{eq10}) in $(tanh(H\,c_{2}-r_{2}),P_{tot}/(\triangle B)^{\beta})$ plane. }
{\label{Fig.4b}}
\end{figure}
Both measurement series consist of 30 points. All measurement points have been taken in temperature $27^{o}C\pm 0.5^{o}C$. Standard deviations per point are equal $12[\frac{W}{m^{3}T^{\beta}}]$ and $15[\frac{W}{m^{3}T^{\beta}}]$ for VITROPERM 500F 18K and SIFERRIT N87, respectively. Applying the formulae (\ref{eq9}) and (\ref{eq10}) and the estimated values of parameters we  have drawn scatter plots which compar estimated points with the experimental ones. Each sample is presented in the two projections.  Note that the estimated model parameters are universal, which means that they work for each $f$ and $\triangle B$. Attention, in order to prevent generation of large numbers in  the estimation process the unit of frequency was kHz while other magnitutes were expressed in SI unit system. 
 \begin{table}[!t]
\caption{Measurement data for SIFERRIT N87}
\label{APPENDIX}
\begin{center}
\begin{tabular}{|c|c|c|c|}\hline
	$\triangle B[T]$	&	$f[kHz]$	&	$H_{DC}[\frac{A}{m}]$	&	$P_{tot}	[\frac{W}{m^{3}}]$\\\hline
	0,201	&	50	&	0,070	&	13155,0	\\
	0,202	&	50	&	4,699	&	13384,6	\\
	0,204	&	50	&	10,70	&	14043,9	\\
	0,201	&	50	&	14,75	&	14338,9	\\
	0,204	&	50	&	19,83	&	15295,3	\\
	0,414	&	50	&	4,920	&	52552,5	\\
	0,411	&	50	&	9,96	    &	54106,4	\\
	0,416	&	50	&	15,11	&	56641,6	\\
	0,411	&	50	&	20,18	&	60464,5	\\
	0,618	&	50	&	4,662	&	118206	\\
   	0,600	&	50	&	10	&	121104\\
    0,619	&	50	&	15,03	&	127222	\\
	0,613	&	50	&	19,85	&	135366	\\
	0,807	&	1	&	4,810	&	582,66	\\
	0,814	&	1	&	10,29	&	582,38	\\
	0,812	&	1	&	15,29	&	618,87	\\
	0,804	&	1	&	19,81	&	682,53	\\
	0,810	&	1	&	25,26	&	829,74	\\
	0,804	&	2	&	0,067	&	1127,81	\\
	0,802	&	2	&	4,780	&	1036,27	\\
	0,805	&	2	&	10,04	&	1010,43	\\
	0,805	&	2	&	14,78	&	1059,07	\\
	0,803	&	2	&	19,81	&	1161,91	\\
	0,808	&	5	&	4,818	&	4066,07	\\
	0,808	&	5	&	15,27	&	4098,72	\\
	0,807	&	5	&	19,59	&	4418,86	\\
	0,806	&	10	&	5,077	&	12875,9	\\
	0,806	&	10	&	10,16	&	12931,8	\\
	0,805	&	10	&	14,78	&	13593,5	\\
	0,806	&	10	&	19,70	&	14891,1	\\\hline
\end{tabular}
\end{center}
\end{table}\vspace{2mm}
\begin{table*}
\begin{center}
\caption{The set of  estimated model's parameters for VITROPERM 500F 18K, formula (\ref{eq10})}
\label{Tab2}
\begin{tabular}{|c|c|c|c|c|c|c|c|c|}\hline
$\alpha$&$\beta$&$x$&$\Gamma_{1}$&$\Gamma_{2}$&$\Gamma_{3}$&$\Gamma_{4}$&$\Gamma_{5}$&$\Gamma_{6}$\\\hline
$\Gamma_{7}$&$y$&$c_{1}$&$c_{2}$&$c_{3}$&$r_{1}$&$r_{2}$&$r_{3}$&-\\\hline
7,848601&	13,40684	&0,3529973&	-101,7297&2390,43&	-334,3942&121,0903&7,88E+00	&1,29E+03\\\hline
-7,66917	&9,77E-01&	4,668659&9,52E-02&	43,98191&0,2228664&	3,042717	&-2,016617&-
\\\hline
\end{tabular}
\end{center}
\end{table*}

\begin{table*}
\caption{The set of  estimated model's parameters for  SIFERRIT N87, formula (\ref{eq9})}
\label{Tab1}
\begin{center}
\begin{tabular}{|c|c|c|c|c|c|c|}\hline
$\alpha$&$\beta$&$x$&$\Gamma_{1}$&$\Gamma_{2}$&$\Gamma_{3}$&$\Gamma_{4}$\\
$\Gamma_{5}$&$\Gamma_{6}$&$\Gamma_{7}$&$\Gamma_{8}$&$c_{1}$&$c_{2}$&$c_{3}$\\\hline
-7,7413	&-11,375&-0,1712&285,292&	118,459&	-27,1697&-2,2167	\\\hline
-5,35E-01&	-3,03E+01	&5,8086	&-5,750&	-0,05529	&8,0579	&0,1102
\\\hline
\end{tabular}
\end{center}
\end{table*}

 \section{Conclusions}
Agreement between the models of core losses ((\ref{eq9}),(\ref{eq10})) and the measurement data confirms the hypothesis that these data obey scaling relations. Now we can construct estimators for core loss in soft magnetic materials exposed to nonsinusoidal periodic
flux waveforms and DC Bias condition. Moreover, thanking to data collapse the space of independent variables is reduced from the three-dimensional one spaned by $[f,\triangle B, H_{DC}]$ to the two-dimensional one spaned by $[\frac{f}{(\triangle B)^{\alpha}}$,$tanh(H_{DC}\cdot c_{i})]$. The consequence which results from this reduction is very significant: both models (\ref{eq9}) and (\ref{eq10}) describe the core loss problem completely (for each $f$, for each $\triangle B$ and for each $H_{DC}$). The question is then how many model parameters must one introduce in order to completely describe the core loss in the 3-dimensional space.

\bibliographystyle{plainnat}

\begin{thebibliography}{99}
\bibitem{bib:Steinmetz}
C.P. Steinmetz, On the law of hysteresis, Trans. Amer. Inst. Elect. Eng., \textbf {9} , 3-64 (1892).
\bibitem{bib:Albach}
M. Albach, T. Durbaum, and A. Brockmeyer, "Calculating core losses in transformers for arbitrary magnetizing currents—a
comparison of different approaches.", IEEE Power Electronics Specialists Conference, 1996, pp. 1463–8.
\bibitem{bib:Reinert}
J. Reinert,  A. Brockmeyer J. Reinert, A. Brockmeyer, and R.W. De Doncker, "Calculation of losses in ferro- and ferrimagnetic materials based on the
modified Steinmetz equation", Annual Meeting of the IEEE Industry Applications Society, 1999.
\bibitem{bib:Li}
Jieli Li, T. Abdallah, and C. R. Sullivan, "Improved calculation of core loss with nonsinusoidal waveforms", in Annual Meeting
of the IEEE Industry Applications Society, 2001, pp. 2203–2210.
\bibitem{bib:Venka}
K. Venkatachalam, C. R. Sullivan, T. Abdallah, and H. Tacca, "Accurate prediction of ferrite core loss with nonsinusoidal
waveforms using only Steinmetz parameters" IEEE Workshop on Computers in Power Electronics (COMPEL), 2002.
\bibitem{bib:Boss1}
Alex Van den Bossche, Vencislav Valchev, Georgi Georgiev, "Measurement and loss model of ferrites in non-sinusoidal
waves", IEEE Power Electronics Specialists Conference, 2004.
\bibitem{bib:Ecklebe}
Jonas M\"{u}hlethaler,  J\"{u}rgen Biela, Johann Walter Kolar and Andreas Ecklebe, Improved Core-Loss Calculation for Magnetic Components Employed in Power Electronic Systems, IEEE TRANSACTIONS ON POWER ELECTRONICS, \textbf{27}, pp.964-973 (2012).
\bibitem{bib:Ecklebe1}
J. M\"{u}hlethaler, J. Biela, J.W. Kolar, A. Ecklebe, "Core Losses Under the DC Bias Condition Based on Steinmetz Parameters,"
IEEE Transactions on Power Electronics, \textbf{27},  pp.953-963, (2012). 
\bibitem{bib:Stan1} 
H.E.~Stanley, "Introduction to Phase Transitions and Critical Phenomena", Oxford: Clarendon Press 1971.
\bibitem{bib:Stan2}
 H.E. Stanley, Scaling, universality, and renormalization: Three pillars of modern critical phenomena, Rev. Mod. Phys.,\textbf {71}, S358-S366 (1999).
\bibitem{bib:Sokal1}
 K.~Sokalski, J.~Szczyg{\l}owski, M.~Najgebauer and W.~Wilczy\'nski, Thermodynamical Scaling of Eddy Current Losses in Magnetic Materials, Proc. 12th IGTE Symposium, ( 2006) pp. 83-86.
\bibitem{bib:Sokal2}
 K.~Sokalski, J.~Szczyg{\l}owski, M.~Najgebauer and W.~Wilczy\'nski, Losses scaling in soft magnetic materials, COMPEL: Int. J. Comput. Math. Electr. Electron. Eng.,\textbf {26}, 640-649 ( 2007).
\bibitem{bib:Sokal3}
 K.~Sokalski, J.~Szczyg{\l}owski,  and W.~Wilczy\'nski, Scaling conception of power loss' separation in soft magnetic materials, arXiv:1111.0939v1 [cond-mat.mtrl-sci] 3 Nov 2011.
\bibitem{bib:Cale}
J. Cale, S.D. Sudhoff, S. D. and R.R. Chan, "A Field-Extrema Hysteresis Loss Model for High-Frequency Ferrimagnetic
Materials," IEEE Transactions on Magnetics,\textbf{44}, issue 7, pp. 1728-1736 (2008).
\bibitem{bib:Boss2}
A. Van den Bossche, et al., Ferrite Losses of Cores with Square Wave Voltage and DC bias, J. Appl. Phys., \textbf{99}, p. 08M908 (2006).
\bibitem{bib:GB2}
 G.~Bertotti, General Properties of Power Losses in Soft Ferromagnetic Materials, IEEE Trans. Magn. \textbf {24}, p.621 (1988).
\bibitem{bib:GB3}
G. Bertotti, F. Fiorillo and G. P. Soardo, THE PREDICTION OF POWER LOSSES IN SOFT MAGNETIC MATERIALS, JOURNAL DE PHYSIQUE, Colloque C8, Supplement au no 12, \textbf{49} (1988).
\bibitem{bib:GB}
 G.~Bertotti, A general statistical approach to the problem of eddy current losses, J. Mag. Mag. Mater. \textbf {41}, p.253 (1984).
\bibitem{bib:Slus}
B. \'{S}lusarek, B. Jankowski, K. Sokalski, J. Szczyg³owski, Characteristics of Power Loss in SMC a Key for Designing the Best 
Values of Technological Parameters, Journal of Alloys and Compounds, \textbf{581}, pp. 699-704 (2013), http://dx.doi.org/10.1016/j.jallcom2013.07.084.
\bibitem{recipes} 
W.H.Press, B.P.Flannery, S.A.Teukolsky, and W.T.Vetterling, {\em Numerical Recipes. The Art of Scientific Computing}, Cambridge Univ. Press 1987, pp. 289-293.

\end{thebibliography}

\end{document}